# Recent advances with a hybrid micro-pattern gas detector operated in low pressure H$_2$ and He, for AT-TPC applications

Marco Cortesi[a], Wolfgang Mittig, Daniel Bazin, Yassid Ayyad Limonge, Saul Beceiro-Novo, Rim Soussi Tanani, Michael Wolff, John Yurkon, and Andreas Stolz

[1] National Superconducting Cyclotron Laboratory, Michigan State University, East Lansing, Michigan 48824, U.S.A

**Abstract.** In view of a possible application as a charge-particle track readout for an Active-Target Time Projection Chamber (AT-TPC), the operational properties and performances of a hybrid Micro-Pattern Gaseous Detector (MPGD) were investigated in pure low-pressure Hydrogen (H$_2$) and Helium (He). The detector consists of a MICROMesh GAseous Structure (MICROMEGAS) coupled to a single- or multi-cascade THick Gaseous Electron Multiplier (THGEM) as a pre-amplification stage. This study reports the effective gain dependence of the hybrid-MPGD at relevant pressure (in the range of 200-760 torr) for different detector arrangements. The results of this work are relevant in the field of avalanche mechanism in low-pressure, low-mass noble gases, in particularly for applications of MPGD end-cap readout for active-target Time Projection Chambers (TPC) in the field of nuclear physics and nuclear astrophysics.

## 1 Introduction

In the field of low-energy nuclear physics, Active-Target Time Projection Chambers (AT-TPCs) have been developed into unique devices to study reactions induced by short-live radioactive isotopes [1], and to break the limits imposed by conventional methods of investigation based on high-resolution spectrometer [2]. AT-TPCs provide several significant features, including 4$\pi$ acceptance of the reaction products; high effective thickness; full detection efficiency and high sensitivity to very low energy events; and an event-by-event reconstruction in three dimensions.

In its basic design, the AT-TPC consists of a large, cylindrical volume of gas, surrounded by a field-shaping cage and terminated at one end by a position-sensitive end-cap chamber. The type of gas that fills the detector volume and acts simultaneously as target and tracking medium, depends upon the requirements of the particular physics case of interest. Most relevant gases are hydrogen (H$_2$) as proton target, deuterium (D$_2$) as deuteron target, helium-4 ($^4$He) as an alpha target, or helium-3 ($^3$He) for studies involving elastic and inelastic scattering, as well as transfer and charge exchange reactions. For investigation of inverse kinematic reactions with TPCs in active target mode, two operative conditions are important:
-) a pure target, specifically no quencher added to the noble gas in order to suppress background reactions.
-) a tuneable pressure of the filling gas (down to a few tens of torr), so that the thickness of the target is adjusted according to the kinematics of the reaction under study.

However, the operation of the position-sensitive endcap detector in low-pressure, pure noble gases is rather difficult: in these conditions, the charge amplification is strongly limited by photo-mediated secondary effects (gas photoionization, feedback loops, etc.) that cause early transition from the proportional avalanche to the streamer mode, and thus leads to spark breakdown [3,4]. This results in a general degradation of the performance of the detector in terms of energy and spatial resolution, counting rate capability, and gain stability.

In view of a possible application as a high-gain, charged-particle track readout for AT-TPC applications, the present work reports recent results of the R&D program carried out at the National Superconducting Cyclotron Laboratory (NSCL), Michigan State University, on the development of a Hybrid Micro-Pattern Gas Detector (MPGD) operated in low pressure low-mass noble gas.

## 2 AT-TPCs at NSCL

At NSCL, two AT-TPCs of new generation have been developed for accurate studying of low-energy nuclear reactions with rare isotopes beams:
-) a half-scale prototype (pAT-TPC) [5], with a volume of 25 litres (50 cm long, 25 cm diameter), equipped with a pad-plane segmented into more than 250 pads. The geometry of anode pad readout can be designed according the type of experiment. The pAT-TPC, being a compact, versatile, and transportable device, has been used to validate the detector concept in first feasibility studies.

---
[a] Corresponding author: cortesi@nscl.msu.edu

-) a full-scale AT-TPC of 200 litres (100 cm long, 50 cm diameter), with an anode plane of more than 10,000 pads read out by the General Electronics for TPC (GET [6]); this device is stationed inside a large bore solenoid magnet which applies a longitudinal magnetic field, up to 2T [1].

Both detectors are equipped with an endcap based on Micromegas [7], which serves as electron avalanche multiplier and tracking device of the reaction products released in the AT-TPC effective volume. Micromegas is unquestionably one of the most successful MPGD structures, providing an excellent position resolution (down to a few tens of μm [8]) and an excellent charge resolution [8-9]. A series of successful experiments have been conducted with the pAT-TPC, including studies of sub-barrier fusion reaction in $^{10}$Be (with Ar/10%CH$_4$ as the gas target) and resonant scattering to investigate α-cluster states in $^{10}$Be [10] and $^{14}$C [11] (with He/20%CO$_2$ as gas target). However, Micromegas is a single-stage, open-geometry multiplier structure and suffers from photon-induced secondary feedback effects, which limit the gas gain when operated without sufficient quench gas.

Figure 1 shows examples of Micromegas gain curves measured in He at different pressures (760, 500 and 300 torr) performed on a standard Micromegas detector (see section 3). A maximum achievable gain of around 30 was measured at 760 torr, while for lower pressure the value decreases to less than 10 for pressure below 500 torr.

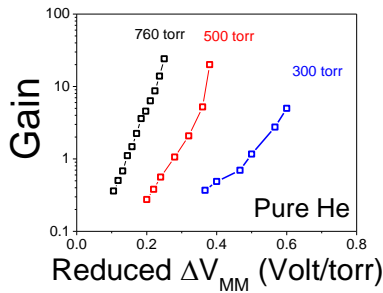

**Figure 1.** Effective gain measured with a Micromegas detector in He at different pressures (760, 500 and 300 torr).

In order to achieve a suitable gas gain during operation with pure low-mass noble gas, the Micromegas readout was combined with pre-amplification stages based on the hole-type THGEM multiplier [14]. This hybrid configuration allow to cascade many multiplier elements and to obtain an overall higher effective gain by manifold electron avalanche processes. Furthermore, THGEM-based detectors permit stable operation at high gain in pure noble gas at low pressure due to [13-15]:

-) the extended dimension of the THGEM holes, typically several times larger than the electron mean-free path even at low pressure;
-) the confinement of the avalanche within the holes, resulting in smaller photon-mediated secondary effects;
-) the quenching effect of small amounts of impurity from natural outgassing of detector components - e.g. N$_2$ acts as wavelength shifter suppressing UV-photons emitted during the avalanche.

Maximum effective gains of $10^4$–$10^7$ have been recently reported [14] with multi-THGEMs structures in pure Helium at pressure ranging from 100 torr up to 760 torr. In the case of H$_2$ and D$_2$, maximum achievable gains of above $10^4$, for pressures of 200 torr and above, were obtained with a double-THGEM detector from single-photoelectrons avalanche [15]. A recent work [16] also reported that the hybrid Micromegas/THGEM detector reduces ion backflow, which in turn results in a substantial increase of the detector gain stability as well as an improved charge resolution.

## 3 Effective gain in He and H$_2$

In this work, basic performance evaluation studies have been carried out using a hybrid (Micromegas/THGEM) detector prototype of an effective area of 10x10 cm$^2$ (Figure 2). The Micromegas structure consisted on a micromesh (18 μm thick stainless steel wire woven at a pitch of 63 μm) held at a distance of 128 μm from the anode plane. The THGEM electrodes are economically produced by mechanically drilling of 0.5 mm diameter holes, spaced by 1 mm (pitch) in a 0.6 mm thick printed circuit board (PCB), followed by Cu etching of the hole's rim (0.1 mm).

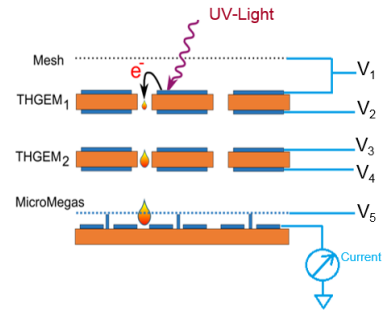

**Figure 2.** Schematic drawing of the detector prototype, consisting of a Micromegas structure coupled to two-cascade THGEM electrodes.

Figure 3 depicts the effective gain as a function of the reduced bias applied symmetrically to the two THGEMs, while the voltage difference across the Micromegas was kept at a constant value of 180 Volt. The measurements were performed in pure He at different pressures (from 760 to 200 torr), and carried out by illuminating the bare Cu-cladded top surface of the first-cascade THGEM with UV-light. The maximum achievable gains, defined by the onset of the discharges, was roughly $10^6$, regardless of the pressure of the filling gas. Notice that for pressures below 200 torr the frequency of discharge increases considerably and no stable operation was possible at high gain.

Following the encouraging results obtained in He with the small-area, test-bench detector [14-15], two large THGEM electrodes (270 mm diameter) have been commissioned and assembled on the pAT-TPC endcap plane (Figure 4).

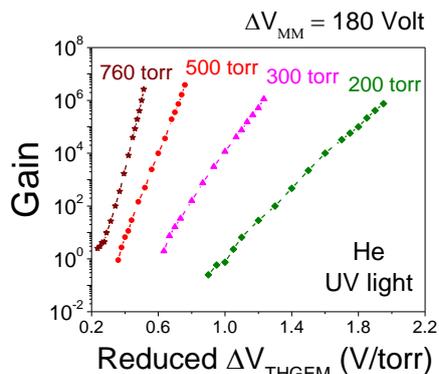

**Figure 3.** Effective gain curves as a function of the reduced bias.

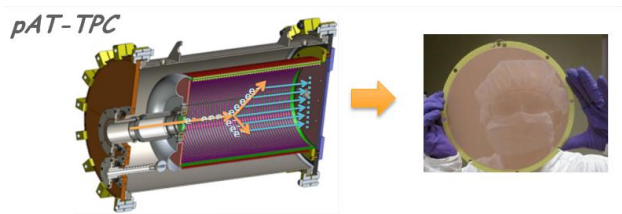

**Figure 4.** Schematic drawing of the prototype detector (right) and photograph of one of the THGEM electrodes (250 mm diameter) mounted in front of the Micromegas pad-plane of the pAT-TPC (left), as charge pre-amplification stage.

The performance evaluation and characterization of the new hybrid endcap detector have been performed in $H_2$, at a pressure of 200 and 300 torr (Figure 5 and 6, respectively). The detector was irradiated with a collimated fission source (252-Cf), mounted on the cathode plane of the pAT-TPC. At these pressures, all of the heavy fragments from the 252-Cf spontaneous fission were stopped inside of the gas volume; the total kinetic energy of the fragments is around 180 MeV.

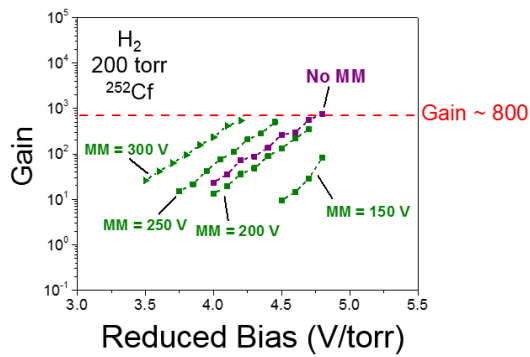

**Figure 5.** Effective gain curves in pure $H_2$ at 200 torr.

In Figure 5 and 6, all the gain curves were obtained by varying the voltage bias applied symmetrically to the two THGEMs (in the graphs expressed as reduced bias – Volt/torr), while the voltage across the Micromegas structure was kept at a constant value (indicated as MM in the figures). The transfer field between the various multiplier stages was kept at a value of 250 Volt/cm. The gain curves of the hybrid detector (green graphs) are compared to the effective gain measured with only THGEM elements as amplification stage (purple graph); in this case the pulse-heights were recorded on the Micromegas mesh. Independently from the voltage bias applied to the Micromegas, upper limits for the maximum achievable gain was achieved at values of 800 and 300, for pressure of 200 and 300 torr respectively. Notice that, for low biases applied to the Micromegas, the significant loss of gain is the result of a low electron transparency of the Micromegas mesh –a substantial number of electrons coming from the THGEM pre-amplification stage are collected on the mesh, instead of being transferred to the anode readout.

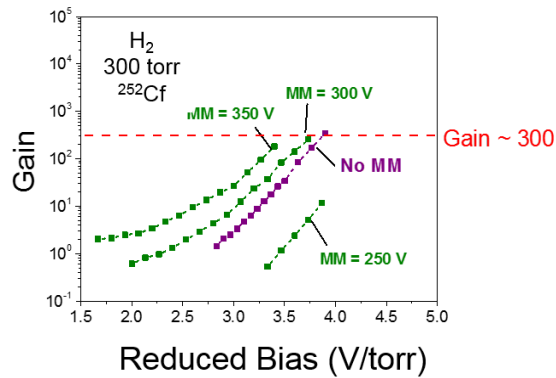

**Figure 6.** Effective gain curves in pure $H_2$ at 300 torr.

As a typical diatomic molecules, $H_2$ (as well as $D_2$) is characterized by many rotational and vibration modes that can be excited by electron impact [15]. These radiationless processes have high cross sections that extend beyond the energy threshold of the ionization process, with the result that an extremely high electric field strength is necessary for a substantial gas avalanche multiplication. Nevertheless, higher voltage biases applied to the various detector elements induced a higher probability of sporadic discharges, resulting in a reduced maximum achievable gain compared to operation in other noble gases.

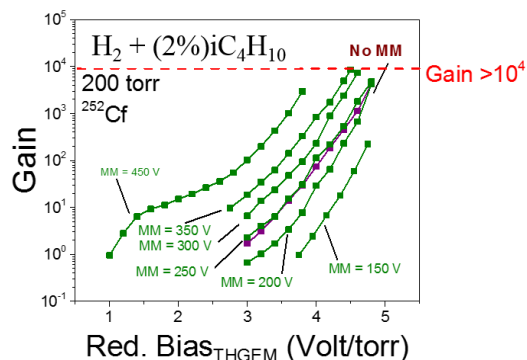

**Figure 7.** Effective gain curves in $H_2/(2\%)_iC_4H_{10}$ at 200 torr.

If higher gain operation is necessary, a small percentage of a quench gas (isobutane or other complex gas molecule, e.g. $CH_4$, $CO_2$, etc.) can be added to the primary noble gas component - $H_2$ (or He). Spark-free operation with an order of magnitude higher effective gain can be achieved, though at the expense of a lower reaction (detection) efficiency and a more complex data analysis - reactions induced with Carbon (or other elements, e.g. Oxygen) need to be identified and rejected. Figure 7 illustrate examples of gain curves measured in $H_2$ with 2% Isobutane admixture at 200 torr.

## 4 AT-TPC calibration

The tracking calibration of the AT-TPC, performed by online monitoring of operating and ambient parameters, is among the various required procedures for keeping a strict tolerance on all the possible sources of inaccuracy. The main goal is to provide the precise information needed for the reconstruction of the particle tracks with sufficient precision, so that the design performance can be achieved.

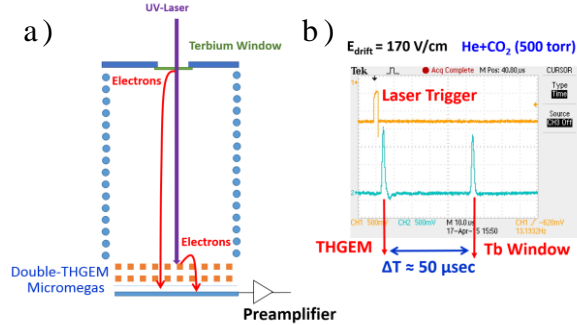

**Figure 8.** Part a): operational principle of the pAT-TPC drift-velocity monitor system (see text for details). Part b): example of recorded photoelectrons pulse-heights signals induced by the pulse UV-light on the pAT-TPC.

The calibration of the drift velocity, which in conjunction with the drift time provides the z-position of the traversing particles, is essential. Variations in drift velocity during operation can be caused by changes of the filling gas composition (due to outgassing of the detector components and/or the gas handling system), as well as temperature, pressure, electric (and magnetic) field fluctuations. The simple approach for determining the drift velocity is by measuring the difference of the arrival time of electrons emitted from two known distances (so-called pick-up electrodes), inside the AT-TPC volume.

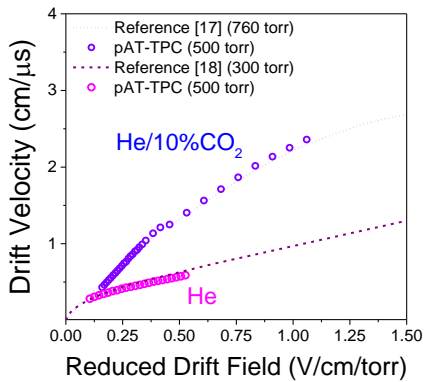

**Figure 9.** Drift velocity in He and He/$CO_2$ as function of the reduced drift field. Data from this work are compared to data available in literature.

The monitor system, which will be implemented on the pAT-TPC, consists of a terbium-coated quartz windows mounted on the cathode plane and illuminated by an intense, pulsed nitrogen laser (337 nm wavelength). Photoelectrons are extracted from the 150 Å thick terbium layer and drifted towards the anode. Part of the UV-light is transmitted through the quartz windows and illuminates the Cu-cladded top surface of the first-cascade THGEM. A copious number of photoelectrons are extracted from the copper (Figure 8b); part of them are focused inside the THGEM holes and then multiplied. Thus each UV-pulse generates two signals (Figure 8b), one due to photoelectrons emitted at the cathode and one due to photo-electrons emitted at the anode plane. The time difference between the two signals (Figure 8b) corresponds to the drift-time of electrons across the entire AT-TPC length (50 cm).

Figure 9 shows the drift velocity as function of the reduced drift field, measured in He and He/$CO_2$ with pAT-TPC monitor system prototype described above. Experimental data from this work are in good agreement with values reported in the literature [17-18].

## 5 Summary

AT-TPCs are a powerful tool for exploiting recent major progress in accelerator technology to perform accurate investigation of non-perturbative quantum systems, by testing weakly bound nuclear species using simple reaction mechanisms. In this work we report recent progress on the development of a hybrid MPGD for AT-TPC applications, consisting of Micromegas coupled to double-cascade THGEMs as pre-amplification stage. The hybrid readout plane detector provides relative high gas gain in pure noble gas (with no gas quencher) at low pressure. This condition is particularly important for beams composed of rare isotopes far from stability, whose intensities are inherently low but their investigations are of prime scientific interest. It is also crucial for reaction studies in the astrophysical domain, where the cross sections are small and an unambiguous event-by-event kinematic reconstruction is essential for identifying particular reaction channels with high sensitivity.